\documentclass{nature}

\usepackage{graphicx}
\usepackage{longtable}

\title{Spectroscopic identification of r-process nucleosynthesis in a double neutron star merger}

\author{E. Pian$^{1}$, P. D'Avanzo$^{2}$, 
S. Benetti$^{3}$,
M. Branchesi$^{4,5}$,
E. Brocato$^{6}$,
S. Campana$^2$, 
E. Cappellaro$^{3}$,
S. Covino$^2$, 
V. D'Elia$^{6,7}$, 
J. P. U. Fynbo$^{8}$,
F. Getman$^9$,
G. Ghirlanda$^2$,
G. Ghisellini$^2$,
A. Grado$^9$,
G. Greco$^{10,11}$,
J. Hjorth$^{8}$,
C. Kouveliotou$^{12}$,
A. Levan$^{13}$,
L. Limatola$^9$,
D. Malesani$^{8}$,
P. A. Mazzali$^{14,15}$,
A. Melandri$^2$, 
P. M{\o}ller$^{16}$,
L. Nicastro$^1$, 
E. Palazzi$^1$, 
S. Piranomonte$^6$, 
A. Rossi$^1$,  
O. S. Salafia$^{17,2}$,
J. Selsing$^{8}$,
G. Stratta$^{10,11}$, 
M. Tanaka$^{18}$,
N. R. Tanvir$^{19}$,
L. Tomasella$^3$, 
D. Watson$^{8}$,
S. Yang$^{20,21}$,
L. Amati$^{1}$,
L. A. Antonelli$^{6}$,
S. Ascenzi$^{6,22,23}$,
M. G. Bernardini$^{24,2}$,
M. {Bo{\"e}r}$^{25}$,
F. Bufano$^{26}$,
A. Bulgarelli$^{1}$,
M. Capaccioli$^{9,27}$,
P. G. Casella$^{6}$,
A. J. Castro-Tirado$^{28}$,
E. Chassande-Mottin$^{29}$,
R. Ciolfi$^{3,30}$,
C. M. Copperwheat$^{14}$,
M. Dadina$^{1}$,
G. De Cesare$^{1}$,
A. Di Paola$^{6}$,
Y. Z. Fan$^{31}$,
B. Gendre$^{32}$,
G. Giuffrida$^{6}$,
A. Giunta$^{6}$,
L. K. Hunt$^{33}$,
G. Israel$^{6}$,
Z.-P. Jin$^{31}$,
M. Kasliwal$^{34}$,
S. Klose$^{35}$,
M. Lisi$^{6}$,
F. Longo$^{36}$,
E. Maiorano$^{1}$,
M. Mapelli$^{3,37}$,
N. Masetti$^{1,38}$,
L. Nava$^{2,39}$,
B. Patricelli$^{40}$,
D. Perley$^{14}$,
A. Pescalli$^{41,2}$,
T. Piran$^{42}$,
A. Possenti$^{43}$,
L. Pulone$^{6}$,
M. Razzano$^{40}$,
R. Salvaterra$^{44}$,
P. Schipani$^{9}$,
M. Spera$^{3}$,
A. Stamerra$^{40,45}$,
L. Stella$^6$,
G. Tagliaferri$^{2}$,
V. Testa$^{6}$,
E. Troja$^{46}$,
M. Turatto$^{3}$,
S. D. Vergani$^{47,2}$,
D. Vergani$^1$
}

\begin{document}
\maketitle

\begin{affiliations}
 \item INAF, Institute of Space Astrophysics and Cosmic Physics, Via Gobetti 101, I-40129 Bologna, Italy
 \item INAF, Osservatorio Astronomico di Brera, Via E. Bianchi 46, I-23807 Merate (LC), Italy
 \item INAF, Osservatorio Astronomico di Padova, Vicolo dell'Osservatorio 5, I-35122 Padova, Italy
 \item Gran Sasso Science Institute, Viale F. Crispi 7, L'Aquila, Italy
 \item INFN, Laboratori Nazionali del Gran Sasso, I-67100, L'Aquila, Italy 
 \item INAF, Osservatorio Astronomico di Roma, Via di Frascati, 33, I-00078 Monteporzio Catone, Italy
\item Space Science Data Center, ASI, Via del Politecnico, s.n.c., 00133, Roma, Italy
 \item Dark Cosmology Centre, Niels Bohr Institute, University of Copenhagen, Juliane Maries Vej 30, DK-2100 Copenhagen {\O}, Denmark
 \item INAF, Osservatorio Astronomico di Capodimonte, salita Moiariello 16, I-80131, Napoli, Italy
 \item Universit\`a degli Studi di Urbino `Carlo Bo', Dipartimento di Scienze Pure e Applicate, P.za Repubblica 13, I-61029, Urbino, Italy
 \item INFN, Sezione di Firenze, I-50019 Sesto Fiorentino, Firenze, Italy
 \item Department of Physics, The George Washington University, Corcoran Hall, Washington, DC 20052, USA
 \item Department of Physics, University of Warwick, Gibbet Hill Road, Coventry CV4 7AL, UK
 \item Astrophysics Research Institute, Liverpool John Moores University, Liverpool Science Park, IC2, 146 Brownlow Hill, Liverpool L3 5RF, UK
\item Max-Planck-Institut f\"ur Astrophysik, Karl-Schwarzschild-Str. 1, 85748 Garching bei M\"unchen, Germany
 \item  European Southern Observatory, Karl-Schwarzschild-Strasse 2, D-85748 Garching bei M{\"u}nchen,  Germany
 \item Dipartimento di Fisica 'G. Occhialini', Universit{\`a} degli Studi di Milano-Bicocca, P.za della Scienza 3, I-20126 Milano, Italy
 \item National Astronomical Observatory of Japan, Mitaka, Tokyo, Japan
 \item Department of Physics and Astronomy, University of Leicester, University Road, Leicester LE1 7RH, UK
 \item Department of Astronomy and Physics, Padova University, Italy
 \item Department of Astronomy, University of California, Davis, USA
\item Dip. di Fisica, Universita` di Roma La Sapienza, P.le A. Moro, 2, I-00185 Rome, Italy
\item Universit\`a di Roma Tor Vergata, Via della Ricerca Scientifica 1, I-00133 Roma, Italy
 \item Laboratoire Univers et Particules de Montpellier, Universit{\'e} Montpellier, CNRS/IN2P3, Montpellier, France
 \item ARTEMIS (UCA, CNRS, OCA), Boulevard de l'Observatoire, CS 34229, F-06304 Nice Cedex 4, France
 \item INAF - Osservatorio Astronomico di Catania, Via S.Sofia 78, I-95123, Catania, Italy
\item Department of physics, University of Naples Federico II, Corso Umberto I, 40, 80138 Napoli, Italy
 \item Instituto de Astrofisica de Andalucia (CSIC), Glorieta de la Astronomia s/n, E-18008 Granada, Spain
\item APC, Universit{\'e} Paris Diderot, CNRS/IN2P3, CEA/Irfu, Obs de Paris, Sorbonne Paris Cit{\'e}, France
\item INFN-TIFPA, Trento Institute for Fundamental Physics and Applications, Via Sommarive 14, I-38123 Trento, Italy
 \item Key Laboratory of dark Matter and Space Astronomy, Purple Mountain Observatory, Chinese Academy of Science, Nanjing 210008, China
 \item University of Virgin Islands, 2 John Brewer's Bay, St Thomas, VI 00802, USA
  \item INAF - Osservatorio Astrofisico di Arcetri, Largo Enrico Fermi 5, I-50125, Florence, Italy
 \item Division of Physics, Mathematics and Astronomy, California Institute of Technology, Pasadena, CA 91125, USA
\item Th{\"u}ringer Landessternwarte Tautenburg, Sternwarte 5, D-07778 Tautenburg, Germany
\item University of Trieste and INFN Trieste, I-34127 Trieste, Italy
\item Institute for Astrophysics and Particle Physics, University of Innsbruck, Technikerstrasse 25/8, A--6020 Innsbruck, Austria
\item Departamento de Ciencias F{\i}sicas, Universidad Andr{\'e}s Bello, Fern{\'a}ndez Concha 700, Las Condes, Santiago, Chile
\item INAF, Osservatorio Astronomico di Trieste, Via G.B. Tiepolo 11, I-34143 Trieste, Italy
\item Scuola Normale Superiore, Piazza dei Cavalieri 7, I-56126 Pisa, Italy
\item Universit{\`a} degli Studi dell'Insubria, via Valleggio 11, I-22100, Como, Italy
 \item Racah Institute of Physics, The Hebrew University of Jerusalem, Jerusalem 91904, Israel
\item INAF, Osservatorio Astronomico di Cagliari, Via della Scienza 5, I-09047 Selargius (CA), Italy
\item {INAF, Istituto di Astrofisica Spaziale e Fisica Cosmica di Milano, via E. Bassini 15, I-20133 Milano, Italy}
\item INAF, Osservatorio Astronomico di Torino, Pino Torinese, Italy
\item NASA, Goddard Space Flight Center, Greenbelt, MD 20771, USA
\item GEPI, Observatoire de Paris, PSL Research University, CNRS, Place Jules Janssen, 92190, Meudon, France

\end{affiliations}


\begin{abstract}
The merger of two neutron stars is predicted to give rise to three major detectable phenomena: a short burst of $\gamma$-rays, a gravitational wave signal, and a transient optical/near-infrared source powered by the synthesis of large amounts of very heavy elements via rapid neutron capture (the $r$-process)\cite{lattimer1977,eichler1989,lipacz1998}. Such transients, named ``macronovae'' or ``kilonovae'' (refs 4-7), are believed to be centres of production of rare elements such as gold and platinum\cite{metzger2017}.  The most compelling evidence so far for a kilonova was a very faint  near-infrared rebrightening in the afterglow of a short $\gamma$-ray burst\cite{tanvir2013,berger2013} at $z = 0.356$, although findings indicating bluer events have been reported\cite{jin2016}.
Here we report the spectral identification and describe the physical properties of a bright kilonova associated with the gravitational wave source GW\,170817\cite{lvcpaper} and $\gamma$-ray burst GRB\,170817A\cite{fermipaper,integralpaper} associated with a galaxy at a distance of 40\,Mpc from Earth.
Using a series of spectra from ground-based observatories covering the wavelength range from the ultraviolet to the near-infrared, we find that the kilonova is characterized by rapidly expanding ejecta with spectral features similar to those predicted by current  models\cite{kasen2015,tanaka2017}. 
The ejecta is optically thick early on, with a velocity of about $0.2$ times light speed, and reaches a radius of $\sim50$ astronomical units in only 1.5 days.
As the ejecta expands,  broad absorption-like lines appear on the spectral continuum indicating atomic species produced by nucleosynthesis that occurs in the post-merger fast-moving dynamical ejecta and  in two slower (0.05 times light speed) wind regions.  Comparison with spectral models suggests that the merger ejected 0.03--0.05 solar masses of material, including high-opacity lanthanides.
\end{abstract}

GW170817 was detected on Aug 17, 12:41:04 UT\cite{lvcpaper}.  A weak short duration ($t \sim 2s$) GRB in the GW error area triggered the Fermi-GBM about two seconds later\cite{fermipaper}, and was detected also by the INTEGRAL SPI-ACS\cite{integralpaper}. 
A significantly improved sky localization was obtained from the joint analysis of LIGO and Virgo data of the GW event, with a 90\% error region of 33.6 square degrees\cite{lvcpaper}. Following this joint GW/GRB detection, a world-wide extensive observational campaign started, using space and ground-based telescopes to scan the sky region were the events were detected. A new point-like optical source (coordinates RA(J2000) = 13:09:48.09, Dec(J2000) = -23:22:53.3)  was soon reported\cite{coulterpaper,valentipaper}, located at 10 arcsec from the center of the S0 galaxy NGC 4993 ($z = 0.00968$\cite{jones2009}) in the ESO 508-G018 group at a distance of 40 Mpc from Earth, consistent with the luminosity distance of the GW signal. It was first named ``SSS17a'' and ``DLT17ck'', but here we use the official IAU designation, AT 2017gfo.

We carried out targeted and wide field optical/NIR imaging observations of several bright galaxies within the reconstructed sky localization of the GW signal with the Rapid Eye Mount (REM) telescope and with the ESO VLT Survey Telescope (ESO-VST). This led to the detection of SSS17a in the REM images of the field of NGC 4993 obtained  12.8 hours after the GW/GRB event.  Following the detection of this source, we started an imaging and spectroscopic follow-up campaign at  optical and NIR wavelengths. Imaging  was carried out with the REM, ESO-VST and ESO-VLT telescopes. A series of spectra was obtained with the VLT/X-shooter, covering the wavelength range 3200--24800\,\AA\, with VLT/FORS2, covering 3500--9000\,\AA, and with Gemini-S/GMOS covering 5500-9000 \AA\ (see ref 20 for GMOS reduction and analysis details).
Overall, we observed the source with an almost daily cadence during the period Aug 18 -- Sep 03, 2017  ($\sim$ 0.5--17.5 days after the GW/GRB trigger; details are provided in the Methods section).  We present here the results of the observations carried out until late August 2017.

As described in the following, the analysis and modelling of the spectral characteristics of our dataset, together with their evolution with time,  result in a good match with the expectations for kilonovae, providing the first compelling observational evidence for the existence of such elusive transient sources. 
Details of the observations are provided in the Methods. 

We adopted a foreground Milky-Way extinction of  E($B-V$) = 0.1\,mag and the extinction curve of\cite{cardelli1989}, and used this to correct both magnitudes and spectra (see Methods). The extinction within the host galaxy is negligible, based on the absence of substaintial detection of characteristic narrow absorption features associated with its interstellar medium. The optical light curve resulting from our data is shown in Figure 1 and the sequence of X-shooter, FORS2, and GMOS spectra in Figure 2. Apart from Milky Way foreground lines the spectrum is otherwise devoid of narrow features that could indicate association with NGC\,4993. In the slit, displaced from the position of the transient from $3''$--$10''$ (0.6--2.0\,kpc in projection), we detect narrow  emission lines exhibiting noticeable structure, both spatially and in velocity space (receding at 100--250 km/s with respect to the systemic velocity) likely caused by the slit crossing a spiral structure of the galaxy (see Methods).

The first X-shooter spectrum of the transient shows a bright, blue continuum 
across the entire wavelength coverage --  with a maximum at $\sim$6000 \AA\ and total luminosity of $3.2 \times 10^{41}$ erg s$^{-1}$ --  that 
can be fit with a black-body of temperature $5000 \pm 200$\,K, and a spherical 
equivalent radius of $\sim 8\times10^{14}$\,cm.
At a phase of 1.5 days after the GW/GRB trigger, this implies an 
expansion velocity of the ejected material of  $\sim 0.2c$.  The temperature is considerably lower than that inferred from photometric observations about 20  hours earlier ($\sim8000$\,K)\cite{malesanigcn21577}, suggesting  rapid cooling. On top of this overall black-body shape are undulations that may represent very broad absorption features similar to those suggested in merger ejecta simulations\cite{tanaka2017}.  We refrain from  connecting these to expansion velocity as they may be blends of many lines with poorly known properties. 

In the second epoch, one day later, where the spectrum only covers the optical range, the maximum  has moved to longer wavelengths, indicating a rapid cooling.  At the third epoch, when information is again available also at NIR wavelengths, the peak has shifted still to 11000\,\AA, and the overall spectral shape is quite different, indicating that the photosphere is receding, the ejecta are becoming increasingly transparent,  and more lines become visible. 
The NIR part of the spectrum evolves in flux and shape much less rapidly. 
Spectrally broad absorption features are observed ($\Delta \lambda / \lambda \sim 0.1-0.2$).  We exclude that these rapid changes can be compatible with supernova time evolution and are instead consistent with a kilonova (see Methods and Extended Data Figure 2).

Unlike in the case of supernova absorption lines, the identification of kilonova atomic species is not secure.  The neutron-rich environment of the progenitors suggests $r$-process nucleosynthesis as the  mechanism responsible for the elemental composition of the ejecta.  Lacking line identification, various plausible nuclear reaction networks are considered and included in models of radiative transfer of kilonova spectrum formation. 
A fraction of the synthesized atoms are radioactive: 
while decaying they heat the ejecta, which then radiate thermally.  All atomic species present in  the ejecta with their various degrees of excitation and ionization absorb the continuum and cause the formation of lines.  The models aim at reproducing these lines assuming a total explosion energy, a density profile and an ejecta abundance distribution.  In kilonovae it is often envisaged that nucleosynthesis takes place in different regions with different neutron excesses and ejecta velocities, typically a post-merger dynamical ejecta region and a disk-wind region.

Various models predict different components and different synthesized masses.  Tanaka et~al.\ (2017)  presented three models with different electron/proton fractions $Y_e$ (see Methods).  We compare our spectra with a scenario where these three components contribute to the observed spectra (Figure 3): a lanthanide-rich dynamical ejecta region with a proton fraction in the range $Y_e$ = 0.1--0.4 and a velocity of 0.2c (orange in Fig. 3), and two slow (0.05c) wind regions of which one has  $Y_e$  = 0.25 and mixed (lanthanide-free and lanthanide-rich) composition (green) and one has  $Y_e$  = 0.30 and is lanthanide-free (blue).  Each of these spectra falls short of the observed luminosity by a factor of $\sim$2, while for other predictions\cite{tanhot2013,kasen2015} the discrepancy is an order of magnitude. In order to investigate the applicability of the model to the present, more luminous, case we have assumed that the involved ejecta mass is larger.  By decreasing the high $Y_e$ (0.3) wind component to 30\% of the value in the original model, and increasing both the intermediate $Y_e$  (0.25) wind component and the contribution of the dynamical ejecta nucleosynthesis by a factor of 2 we obtain a satisfactory representation of the first spectrum (Figure 3). 

Although direct rescaling of these models is not in principle correct (for larger masses we can expect that the spectrum of each ejecta could change) we can estimate that the ejected mass was $\sim$ 0.03 -- 0.05 M$_\odot$, and that the high $Y_e$ wind ejecta (blue line) are significantly suppressed, possibly because of viewing angle away from the GRB or a narrow jet angle or both. It is also suggestive that a wide range of $Y_e$ values are realised in the ejecta, possibly as a function of latitude.

At successive epochs, the same components represent in a less satisfactory way the observed spectral features, which indicates that the set of adopted opacities is not completely adequate, as the cooling of the gas is not properly followed by lines of different ionization states, and that the radioactive input may also not be accurately known.

Because a short GRB was detected in association with a  GW trigger, we evaluated the expected contribution of its afterglow at the epochs of our observations. Nine days after GW170817 trigger time, an X-ray source  was discovered by Chandra at a position consistent with the kilonova, at a flux level of $\sim 4.5\times 10^{-15}$ erg cm$^{-2}$ s$^{-1}$ (0.3--8 keV).
This source could be delayed X-ray afterglow emission from GRB170817A, produced by an off-beam jet\cite{trojapaper}.  This may account for the otherwise small probability of having an aligned short GRB jet within such a small volume\cite{patricelli2016}. 
The X-ray emission is compatible with different scenarios: a structured jet with an energy per solid angle decreasing with the angular distance from the axis, viewed at large angles  (e.g.\cite{salafia2015}), a cocoon accelerated quasi--isotropically at mildly relativistic velocities by the jet\cite{lazzati2017,nakar2017} or a simple uniform jet observed at large angles. All these scenarios predict an optical afterglow much fainter than the kilonova (see Methods).   On the other hand, if we assume that the early (0.45 days) optical flux we measured is afterglow emission, we estimate, at the same epoch, an X-ray flux $>10^{-12}$ erg cm$^{-2}$ s$^{-1}$ and a 6 GHz radio flux density of $\approx$ 10 mJy. These estimates are not consistent with the absence of X-ray and radio detections at the corresponding epochs\cite{bannister2017,evanspaper}.

Our long and intensive monitoring and wide wavelength coverage enabled the unambiguous detection of time-dependent kilonova emission and sampled  fully its time evolution.  This not only confirms the association of the transient with the GW, but, combined with the short GRB detection, also proves beyond doubt that at least a fraction of short duration GRBs are indeed associated 
with compact star mergers. Furthermore, this first detection provides important insights on the environment of merging NSs. The counterpart's location is only $\sim{}2$ kpc (projected distance) away from the center of an early-type galaxy. This is a quite common offset for short GRBs (e.g.\cite{fong2010}) and is consistent with predictions from theoretical models of merging NSs (e.g.\cite{belczynski2006}). Moreover, the counterpart's location does not appear to coincide with any globular cluster, which suggests a field origin for this NS binary. The nearest possible globular clusters are at $> 2.5^{\prime\prime}$ (corresponding to 500 pc) from the source position\cite{levanpaper}.
The formation channel of this event would be best explored with future modeling and simulations. Finally, since this GRB was rather under-energetic (isotropic gamma-ray output of $\sim 10^{46}$ erg) and likely off-axis with respect to the line of sight, we conclude that  there may be a 
large number of similar nearby off-axis short bursts that are not followed up  at frequencies lower than gamma-rays.  These are also GW emitter candidates and the present event has demonstrated how the search of the randomly oriented parent population of short GRBs can be made effective via coordinated gravitational interferometry and multi-wavelength observations. 
%


\bibliographystyle{naturemag}


%


\begin{addendum}
 \item
	Work in this paper was based on observations made with ESO  Telescopes at the Paranal Observatory under programmes ID  099.D-0382 (PI: E.Pian), 099.D-0622 (PI: P: D'Avanzo), 099.D-0191 (PI: A. Grado) and with the REM telescope at the ESO La Silla Observatory under program ID 35020 (PI: S. Campana). Gemini observatory data were obtained under programme GS-2017B-DD-1 (PI: L. P. Singer). We thank the Gemini Observatory for performing these observations, the ESO Director General for allocating Discretionary Time to this program and the ESO operation staff for excellent support of this program.
We acknowledge INAF for supporting the project ``Gravitational Wave Astronomy with the first detections of adLIGO and adVIRGO experiments - GRAWITA'' PI.: E. Brocato. We acknowledge support from the ASI grant I/004/11/3.
J. Hjorth was supported by a VILLUM FONDEN Investigator grant (project number 16599). M.M.K. acknowledges support from the GROWTH (Global Relay of Observatories Watching Transients Happen) project funded by the National Science Foundation under PIRE grant number 1545949.

\bigskip
\noindent

\item[Author contribution]
E. Pian and P. D'Avanzo are PIs of the two active ESO VLT programs and coordinated the work.  J.~Selsing reduced all the X-shooter spectra presented in Figure 2 and wrote the relevant sections.  M. Tanaka developed the kilonova spectral models. E. Cappellaro assisted with the spectral analysis.  
P. Mazzali  provided the liaison between spectral observations and kilonova theory: he coordinated the  theoretical interpretation,  developed the match between the synthetic and   observed spectra (Figure 3), and wrote the part on their description and discussion. 
S. Campana coordinated the REM observations.  S. Covino, A. Grado and A. Melandri reduced and analysed the optical photometry (Figure 1).  M. Kasliwal provided the Gemini spectrum.  D. Malesani assisted with early observation planning.  G. Ghirlanda, G. Ghisellini and O. S. Salafia wrote the section on the off-beam jet with contributions from L. Amati, Y.Z. Fan, Z.P. Jin, T. Piran, A. Stamerra and B. Patricelli. D. Watson assisted with the analysis of spectra in light of thermal models and assisted with paper writing.
E. Brocato was the Principal Investigator of the GRAvitational Wave Inaf TeAm (GRAWITA)  for GW  electromagnetic follow-up.
M. Branchesi liaised GRAWITA  with LIGO-VIRGO collaborations activities.
A. Grado coordinated the ESO-VST observations.
L. Limatola and F. Getman developed the pipeline to reduce the VST data.
N. Tanvir and A. Levan assisted with NIR data calibration issues.
J. P. U. Fynbo, J. Hjorth and C. Kouveliotou assisted with paper writing and short GRB expertise.
L. Nicastro supervised the data flow and handling.
S. Piranomonte and V. D'Elia contributed to the data reduction and analysis of the X-shooter spectra.
E. Palazzi, A. Rossi, G. Stratta and G. Greco participated in the organization of the observations and image analysis and provided specific input for photometry calibration.
L. Tomasella, S. Yang, and S. Benetti contributed to the data analysis, with particular reference to ISM spectral  features.
P. M\o ller assisted with issues related to ESO policies and observation planning.
This effort was led by  GRAWITA, that includes most co-authors, and is based on GW electromagnetic follow-up programs at ESO and at many telescopes both in Italy and at the Canary Islands. All GRAWITA members contributed to the work development at many phases from preparation of proposals, coordination with the LIGO-VIRGO collaborations, activation of approved programs at many facilities, data acquisition, reduction, analysis, interpretation and presentation.
\bigskip
\noindent

\item[Author Information]
Reprints and permissions information are available at www.nature.com/reprints

\noindent The authors declare that they have no competing financial interests.

\noindent Correspondence and requests for materials should be addressed to E.~Pian (e-mail: pian@iasfbo.inaf.it).

\end{addendum}

\clearpage


\newpage

\methods


\subsection{Optical/NIR imaging}\label{sec:phot}
Our first observations of the field of SSS17a were carried out with the 60-cm robotic telescope REM\cite{chincarini2003} located at the ESO La Silla Observatory (Chile) in the g, r, i, z and H bands starting on 2017 Aug 18 at 01:29:28 UT (i.e. 12.8 hours after the GW event). The field was included in the selection we made to carry out targeted observations of catalogued galaxies in the LVC skymap aimed at searching for an optical/NIR counterpart of the GW event starting on 2017 Aug 17 at 23:11:29 UT (i.e. 10.5 hours after the GW event)\cite{melandrigcn1,melandrigcn2}. Following this first detection, we started an extensive follow-up campaign of optical/NIR imaging carried out with an almost daily cadence from about 1.5 to 15.5 days after the time of the GW trigger. These observations were performed using the ESO VLT telescopes equipped with the  X-shooter acquisition camera, the FORS2 instrument, and the ESO VST equipped with OmegaCam instrument\cite{piangcn1,davanzogcn1,gradogcn1,gradogcn2}. The complete log of our photometric observations is reported in Extended Data Table 1. The optical/NIR light curves are shown in Figure 1. 
Concerning REM and FORS2 imaging, data reduction was carried out following the standard procedures: subtraction of an averaged bias frame and division by a normalized flat frame. The astrometric solution was computed against the USNO-B1.0 catalogue (\texttt{http://www.nofs.navy.mil/data/fchpix/}). 
Aperture photometry was performed using SExtractor\cite{bertin1996} and the PHOTOM package part of the Starlink software distribution (\texttt{http://starlink.eao.hawaii.edu/starlink}). The photometric calibration was achieved by observing  Landolt standard fields and the Pan-STARRS catalogue (\texttt{https://panstarrs.stsci.edu}). In order to minimize any systematic effect, we performed differential photometry with respect to a selection of local isolated and non-saturated reference stars. As shown in Extended Data Figure~1, the transient is embedded in the host galaxy light, so that the background around the transient position is highly inhomogeneous, making accurate photometry measurements arduous. In order to minimize the effect of flux contamination from the host light, we fitted it with an analytical profile. The result obtained from the fit was then subtracted from the image in a neighborhood  of the transient. This  procedure was repeated for each frame. After this subtraction, the background around the transient position is much more uniform, enabling accurate photometric measurements. A dedicated procedure was applied for the reduction and analysis of the wide-field images obtained with the VLT Survey Telescope (VST\cite{capaccioli2011}). The telescope is equipped with OmegaCam \cite{kuijken2011}, a camera with one square degree field of view (FOV) matched by 0.21 arcsec pixels scale.  
Data have been processed with a dedicated pipeline for the VST-OmegaCAM observations (dubbed VST-tube\cite{grado2012}). The pipeline searches for new data in the ESO Data archive and, if available, automatically downloads and processes them  performing the following main steps: pre-reduction; astrometric and photometric calibration; mosaic production. The OT magnitude, in the AB system, is the PSF fitting magnitude measured on the image after subtracting a model of the galaxy obtained fitting the isophotes with the IRAF/STSDAS task ELLIPSE \cite{tody1993}. The reference catalog used for the absolute photometric calibration is the APASS DR9.

\subsection{FORS2 spectroscopic observations}\label{sec:forsred}

FORS2 spectra were acquired with the 600B and 600RI grisms, covering the 3500--8600~\AA{} wavelength range. We used in all cases a $1''$ slit, for an effective resolution of $R \sim 800-1000$. Spectral extraction was performed with the IRAF software package (IRAF is the Image Reduction and Analysis Facility made available to  the  astronomical  community  by  the  National  Optical  Astronomy Observatories, which are operated by AURA, Inc., under contract with the US National Science Foundation. It is available at http://iraf.noao.edu.). Wavelength and flux calibration of the spectra were accomplished using helium-argon lamps and spectrophotometric stars. A check for slit losses was carried out by matching the flux-calibrated spectra to our simultaneous photometry  (see Extended Data Table 1 and Extended Data Table 2). This shows that the derived spectral shape is robust.

\subsection{X-shooter spectroscopic observations}\label{sec:xsred}

The cross-dispersed echelle spectrograph, X-shooter\cite{vernet2011}, mounted on the VLT, was used to observe the optical/near-infrared counterpart of GW170817. The observing campaign
started on the  night following the discovery and continued until the source had faded below the detection limit (see Extended Data Table 2) of X-shooter. 
The observations were carried out using a standard ABBA nodding pattern.
Similar position angles of the slit were used for all observations. The position of the slit on the source is shown in Extended Data Figure 1.

The spectroscopic data obtained with X-shooter were managed with the Reflex interface\cite{freudling2013} and reduced using version 2.9.3 of the X-shooter pipeline\cite{modigliano2010}. The reduction cascade consists of bias subtraction, order tracing, flat fielding, wavelength calibration, flux calibration using the spectrophotometric standard EG274 \cite{moehler2014}, background subtraction and order rectification -- all carried out using the nightly obtained calibration files. A refinement to the wavelength solution was obtained by cross correlating the observed sky spectra with a synthetic sky spectrum\cite{noll2012, jones2013}, leading to a wavelength solution more accurate than 1 km s$^{-1}$. Because X-shooter is a cross-dispersed echelle spectrograph, the individual echelle orders are curved across each detector and a rectification algorithm, which correlates neighboring pixels, must be employed. A sampling of 0.2/0.2/0.6 \AA~per pixel (in the UVB, VIS, and NIR arms, respectively) in the rectified image was chosen to minimize this correlation while conserving the maximal resolving power. The effective resolving power, $R$, of each observation was obtained from fits to unsaturated telluric absorption lines and yielded mean values of 4290/8150/5750 in the UVB/VIS/NIR arms, respectively. This is better than nominal values, owing to a seeing PSF being narrower than the slit width. Immediately following the observations each night, telluric standard stars were observed at an airmass comparable to the target from which the atmospheric transmission spectrum was obtained using Molecfit\cite{smette2015, kausch2015}. 
Host continuum contamination is visible as a faint background gradient along the slit. An effort has been made to minimize this contamination by using the background regions closest to the target. 
The images are combined in nightly sets using a weighting scheme based on a moving background variance measure wide enough to avoid it being pixel based and therefore unsuitable for Poisson-noise dominated images.
For a subset of the observations, the signal-to-noise (S/N) in the spectral trace is large enough to build a model of the spectral line-spread function to employ an optimal extraction algorithm \cite{horne1986}, but for the majority of the data, an aperture covering the entire trace is used. To establish an accurate flux calibration, slit loss corrections were calculated using the average seeing FWHM of the nightly observations along with the theoretical wavelength dependence of seeing \cite{fried1966}. The slit losses are obtained by integrating a synthetic 2D PSF over the width of the slits and corrections are made accordingly.

\subsection{Foreground dust extinction}

We have estimated the intervening dust extinction toward the source using the Na\,I\,D line doublet at 5896\,\AA.  Based on the strength of the line in our Galaxy we derive E($B-V$) = 0.09\,mag using component D1,  E($B-V$) = 0.05\,mag using component D2, and E($B-V$) = 0.06\,mag using the sum\cite{poznanski2012}.  The Galactic extinction is thus limited to E($B-V$) $<$ 0.1\,mag.   Similar upper limits on E($B-V$) are obtained from  the upper limits on the equivalent widths of  the  undetected K\,I~7699\,\AA\ absorption line\cite{munzwi1997}  (EW $< 0.025$ \AA) and undetected 8620 \AA\ diffuse interstellar band\cite{munari2008} (EW $< 0.04$ \AA).  These estimates and limits are marginally consistent with the value of  E($B-V$) = 0.11 mag obtained from COBE/DIRBE maps covering that sky region\cite{schlafly2011}.

\subsection{Spectrum analysis and interpretation}

The first epoch X-shooter spectrum was fit with a black-body with temperature of $5000 \pm 200$ K.
The main deviations from this fit are two absorption-like lines at 8100 and 12300 \AA, that evolve with time and become more pronounced in the second spectrum.  Altogether,  all deviations from a black-body in the first spectrum are below $\sim$10\%  from 3500 \AA\ to 20000 \AA, indicating that the fit is very satisfactory. Moreover,  the expansion speed of $0.2c$ we derive from the black-body radius at the epoch of the first spectrum (1.5 days) is compatible with the width of the absorption lines we observe  in the second spectrum ($\Delta \lambda / \lambda \sim 0.1-0.2$), confirming  that the black-body emission in the first spectrum is highly efficient.

The first 4 X-shooter spectra were compared with kilonova models from Tanaka et al. (2017). The model uses atomic structure calculations for Se (Z = 34), Ru (Z = 44), Te (Z = 52), Ba (Z = 56), Nd (Z = 60), and Er (Z = 68) to construct the atomic data for a wide range of r-process elements. By using two different atomic codes, they confirmed that the atomic structure calculations returned uncertainties in the opacities by a factor of up to $\sim$2.  Thereafter, they apply multiwavelength radiative transfer simulations to predict a possible variety of kilonova emission.   For each model, the abundance is assumed to be homogeneous in the ejecta,  However,  a high-$Y_e$ component should preferentially dominate near the polar region and low-$Y_e$/dynamical component develops  in the equatorial region. For each model, the energy release is similar to a power-law ($t^{-1.3}$) owing to the sum of the radioactive decays of various nuclei with different lifetimes.  The efficiency of the energy deposition is also taken into account, and the energy deposition rate is somewhat steeper than $t^{-1.3}$  because the gamma-rays can escape without depositing energy.

We emphasize that we have not attempted a real fit of this model to our X-shooter spectra, but have rather looked into an interpretation that was in reasonable agreement.  The match is satisfactory only for the first X-shooter spectrum, and not completely satisfactory for the following three.  For this reason, we refrained from deriving a light curve model.  Infact, in principle,  one may fold the  synthetic spectral  model with the sensitivity curve of any given broad-band filter and integrate the flux in the corresponding band to compare with the observed one.  However, the result may be misleading independent of how persuasive it is at face value.  The spectral comparison allows one to appreciate in which wavelength ranges the model is effective and in which ones it fails.  Integration of the model over a broad wavelength interval cancels the spectral "memory" and prevents a critical judgment.  In other words, since the spectral model is not completely satisfactory,  the comparison of synthetic and observed photometry is not significant, although it may appear good.


\subsection{Description of the spectral evolution}\label{sec:specev}

The first X-shooter  spectrum obtained at $t=1.5$ d after the GW trigger shows an almost featureless, moderately blue continuum. The overall spectral energy distribution is similar to that of early, broad line core collapse SNe. 
While in general at this relatively low temperature ($\sim$5000 K) SNe typically show strong broad features using the supernova spectral classification tool GELATO \cite{Harutyunyan} a good match is obtained with the early spectra of the type Ib SN2008D/XRF080109\cite{mazzali2008}.
As shown in Extended Data Figure 2, the X-shooter extended spectral range  displays, by comparison with the black-body fit (dotted line) the presence of some large scale modulations that are suggestive of multi-component contributions already suggestive of a kilonova event.

In the next two days  the spectrum shows a very rapid  evolution. The continuum temperature rapidly drops to about 3300K and broad features emerges, with peaks at 10700 \AA\ and 16000 \AA. The broad features point to very  high expansion velocity and the rapid evolution to a low ejected mass. 
The combined spectral properties and evolution are unlike those of any known SN types and instead they are very similar to the predicted outcomes of kilonova models.

In the following week the temperature derived from the optical continuum seems to remain roughly constant while the peak at 10700 \AA\ drifts to longer wavelengths (11200 \AA\ at day 6) and decreases in intensity until, at 
ten days from discovery, the dominant feature in the spectrum is a broad emission centered at about 21000 \AA.

\subsection{Host emission analysis}\label{sec:hostlines}

Extending 3--10$^{\prime\prime}$ (0.6 -- 2.0\,kpc in projection) from the position of the GW counterpart are emission lines formed in the host. The lines are identified as [O II]$\lambda 3726,3729$, H$\beta$, [O III]$\lambda 4959,5007$, H$\alpha$, [N II]$\lambda 6549,6583$ and [S II]$\lambda 6717,6731$, and they exhibit both spatial and velocity structure along the extent of the slit, as shown in Extended Data Figure 3.  

From the brightest blob of emission, centered at 6$^{\prime\prime}$ (1.2\,kpc in projection) from the source, we measure a receding velocity of $247 \pm 15\,{\rm km}\,{\rm s}^{-1}$ relative to the host nucleus (adopting a systemic velocity of NGC 4993 of
$2916 \pm 15\,{\rm km}\,{\rm s}^{-1}$). Along the spatial direction of the slit, closer to the
source, the emission line centroids become more blue-shifted, approaching a recession velocity of $100\,{\rm km}\,{\rm s}^{-1}$ relative to the NGC 4993 systemic velocity. The velocity range ($150\,{\rm km}\,{\rm s}^{-1}$) of the line emission along the slit indicates coherent motion of the gas along the slit. This is further supported by the dust lanes superposed on the host nucleus\cite{coulterpaper,panpaper}. The presence of spiral arms was also noted by\cite{levangcn1}. A strong [N II]$\lambda$6583 relative to H$\alpha$ combined with a weak H$\beta$ relative to [O III]$\lambda$5007 indicates a radiation field dominated by AGN activity, as also reported previously\cite{kasliwalpaper,hallinanpaper,cookepaper} and supported by the presence of a central radio source\cite{alexanderpaper}. Using the Balmer decrement, the inferred extinction at the position of the line emission is E($B-V$) $= 0.21 \pm 0.21$.

\subsection{Off--beam jet scenario}\label{sec:offj}

GRB170817A had a fluence of $2.2\times 10^{-7}$ erg cm$^{-2}$ in the 10-1000 keV energy range as observed by the GBM which, at a distance of 40 Mpc,  corresponds to a $\gamma$--ray isotropic equivalent energy $E_{\rm iso}\sim4.3\times10^{46}$ erg. The peak energy is $E_{\rm peak}=128\pm48$ keV\cite{fermipaper,goldsteingcn1}. The observed $E_{\rm iso}$ is three to four orders of magnitude smaller than the average energy of short GRBs with known redshift\cite{davanzo2014,berger2014}.

For illustration let us consider a very simple model: a uniform conical jet of semi-aperture angle $\theta_{\rm jet}$ observed off--beam, i.e at a viewing angle $\theta_{\rm view}>\theta_{\rm jet}$.  In this case larger bulk Lorentz factors $\Gamma$ correspond to larger de--beaming factors $b=E_{\rm iso}(0^\circ)/E_{\rm iso}(\theta_{\rm view})$ for a fixed $\theta_{\rm view}$\cite{ghisellini2006,salafia2016}. Given the small distance of 40 Mpc, and a likely luminosity function decreasing with increasing luminosity (e.g. \cite{wanderman2015,ghirlanda2016supplement}), we can assume that the on--axis luminosity of this burst belongs to the low--luminosity tail. For this reason we assume  $E_{\rm iso}(0^\circ)=10^{50}$ erg. Therefore $b=2500$. The probability of a jet oriented at an angle $<\theta_{\rm view}$ is $P(<\theta_{\rm view})=1-\cos\theta_{\rm view}$. A probability of at least $P>10$\% implies $\theta_{\rm view}>26^\circ$. 
An off-axis viewing angle larger than $\sim 30^\circ$ is also suggested by the expected rate of joint GW and Fermi-GBM detection\cite{patricelli2016} rescaled to the actual observations.
Combining Eq. 2 and 3 from\cite{ghisellini2006} it is possible to estimate the observed energy $E_{\rm iso}$ and peak energy $E_{\rm peak}$ as a function of $\theta_{\rm view}$ and $\Gamma$ for a given $\theta_{\rm jet}$. With $\theta_{\rm view}=30^\circ$, $b=2500$ ($E_{\rm iso}(0^\circ)=10^{50}$ erg) requires $\Gamma=10$ for $\theta_{\rm jet}=10^\circ$. The latter is within the currently few estimates of short GRB opening angles\cite{fong2016supplement} and $\Gamma\sim10$ is within the dispersion of the $\Gamma-E_{\rm iso}$ relation\cite{ghirlanda2012,liang2013} for $E_{\rm iso}(0^\circ)\sim 10^{50}$ erg. With these values $E_{\rm peak}(0^\circ)$ turns out to be $\sim$2 MeV. The corresponding comoving frame peak energy would be $\sim$100 keV. If photons with much larger energies are absorbed by pair production we should expect (as observed at $30^\circ$) a spectral cutoff at $\sim$650 keV which is larger than the observed peak energy reported by the GBM. Though these values of $E_{\rm peak}(0^\circ)$ and $E_{\rm iso}(0^\circ)$ are consistent with those observed in short GRBs, they locate this burst relatively far from the possible spectral-energy correlations of short GRBs.

Extended Data Figure 4 shows the predicted afterglow light curves at 6 GHz, $R$ band and 1 keV. The filled circle shows the X--ray flux at 15 days\cite{trojapaper,haggardpaper}. The arrows  show two representative radio upper limits: at 8.65 days (obtained\cite{Paragi} by co-adding six e-MERLIN observations at 5 GHz) and at 20 days (obtained\cite{Mooley} with MeerKAT at 1.5 GHz).  For the model curves the assumed parameters are: $\theta_{\rm jet}=10^\circ$,  $\theta_{\rm view}=30^\circ$, isotropic equivalent kinetic energy $E_{\rm k, iso}=10^{50}$ erg, $\Gamma=10$, a uniform density ISM with $n=2\times10^{-3}$ cm$^{-3}$ and standard micro-physical parameters at the shock i.e. $\epsilon_{\rm e}=0.1$, $\epsilon_{\rm B}=0.01$ and electrons' energy injection power law index $p=2.1$. Standard afterglow dynamics and radiation codes\cite{vaneerten2010} are used. As can be seen the R flux is always below $2\times10^{-5}$ mJy, corresponding to R$>$28, and therefore orders of magnitude lower than the kilonova emission.

\begin{addendum}

 \item[Data Availability:] The data that support the plots within this paper and other findings of this study are available from the corresponding author upon reasonable request.

\end{addendum}

\newpage

%
%
\begin{figure}
	\centering
\includegraphics[height=14.cm,angle=0]{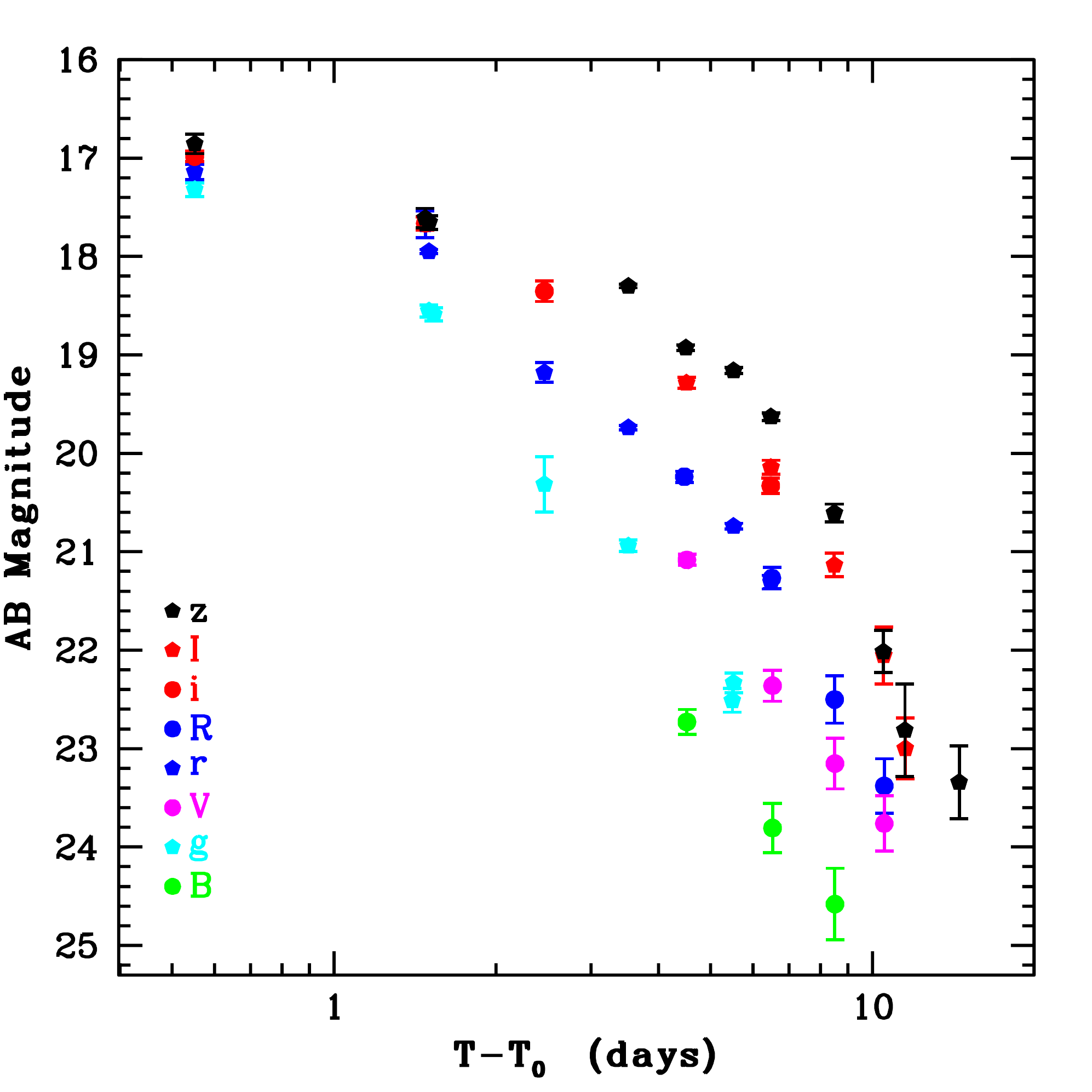}
\label{fig:opticallc}
\end{figure}

\noindent {\bf Figure 1: Multiband optical light curve of AT 2017gfo.} The data shown for each filter (see legend) are listed in Extended Data Table 1. Details of data acquisition and analysis are reported in Methods. The x axis indicates the difference in days between the time at which the observation was carried out $T$ and the time of the gravitation-wave event $T_0$. The error bars show the $1\sigma$ confidence level. The data have not been corrected for Galactic reddening.

%
%
\begin{figure}
	\centering
	\includegraphics[height=14.cm,angle=0]{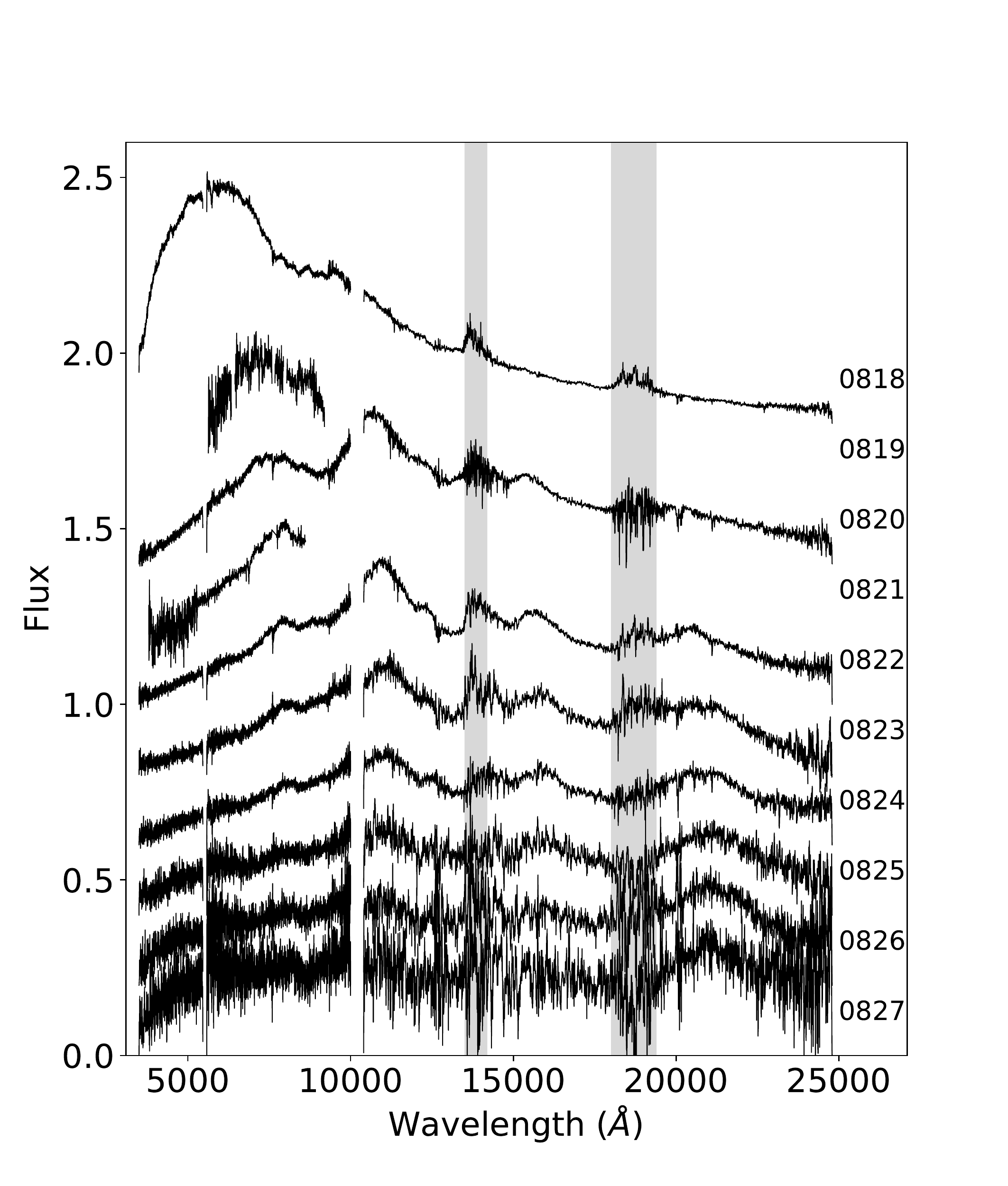}
	\label{fig:obsspectra}
\end{figure}

\noindent {\bf Figure 2: Time evolution of the AT 2017gfo spectra.} VLT/X-shooter, VLT/FORS2 and Gemini/GMOS spectra of AT 2017gfo. Details of data acquisition and analysis are reported in Methods. For each spectrum, the observation epoch is reported on the left (phases with respect to the gravitation-wave trigger time are reported in Extended Data Table 2; the flux normalization is arbitrary). Spikes and spurious features were removed and a filter median of 21 pixels was applied. The shaded areas mark the wavelength ranges with very low atmospheric transmission. The data have not been corrected for Galactic reddening.

%
%
\begin{figure}
	\centering
	\includegraphics[height=17.cm,angle=0]{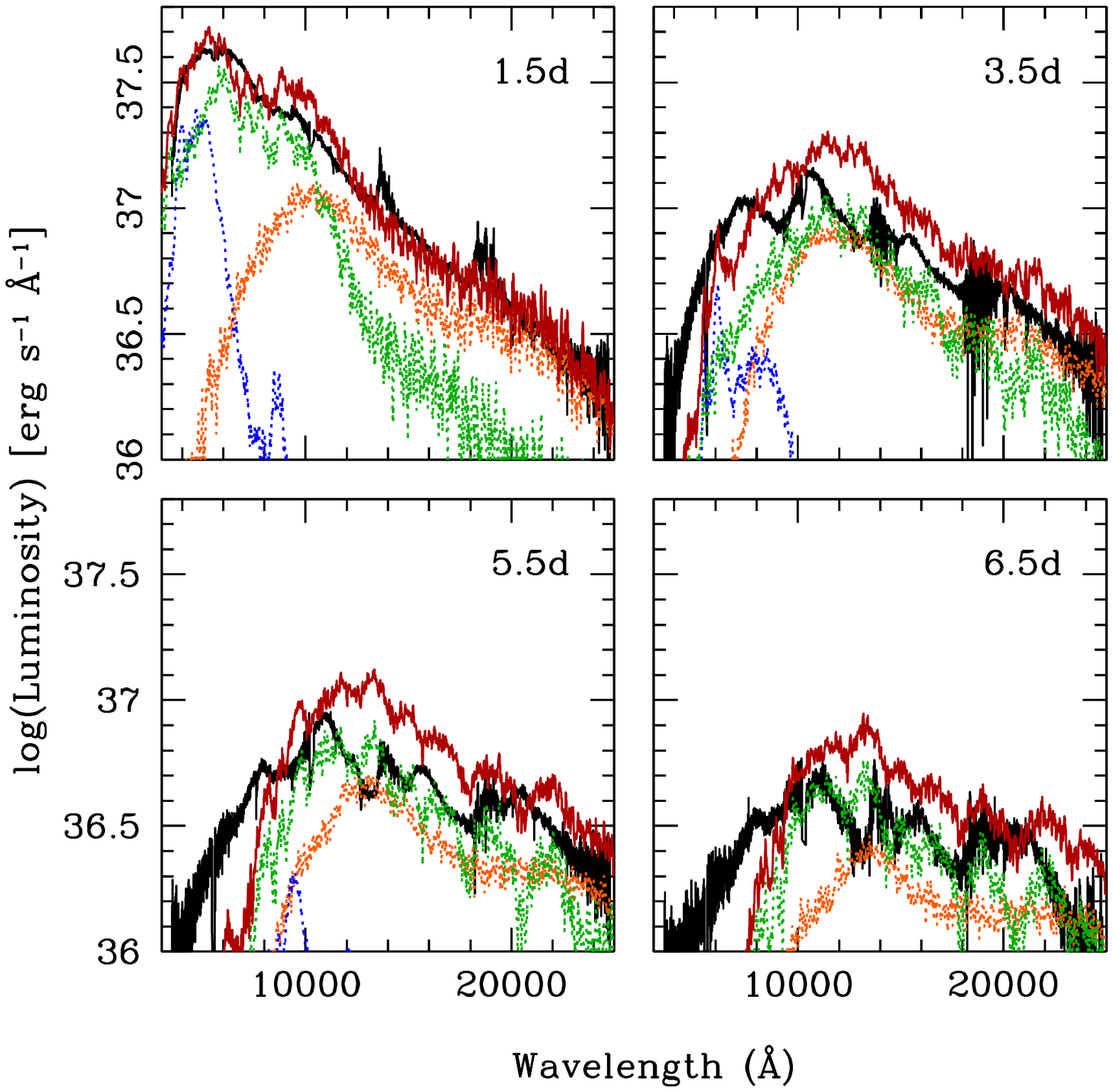}
	\label{fig:spectramodels}
\end{figure}

\noindent {\bf Figure 3: Kilonova model compared to the AT 2017gfo spectra.} X-shooter spectra (black line) at the first four epochs and kilonova models: dynamical ejecta ($Y_e = 0.1-0.4$, orange), wind region with proton fraction $Y_e = 0.3$ (blue) and $Y_e = 0.25$ (green). The red curve represents the sum of the three model components.


%
%
\begin{figure}
	\centering
	\includegraphics[height=18.cm,angle=0]{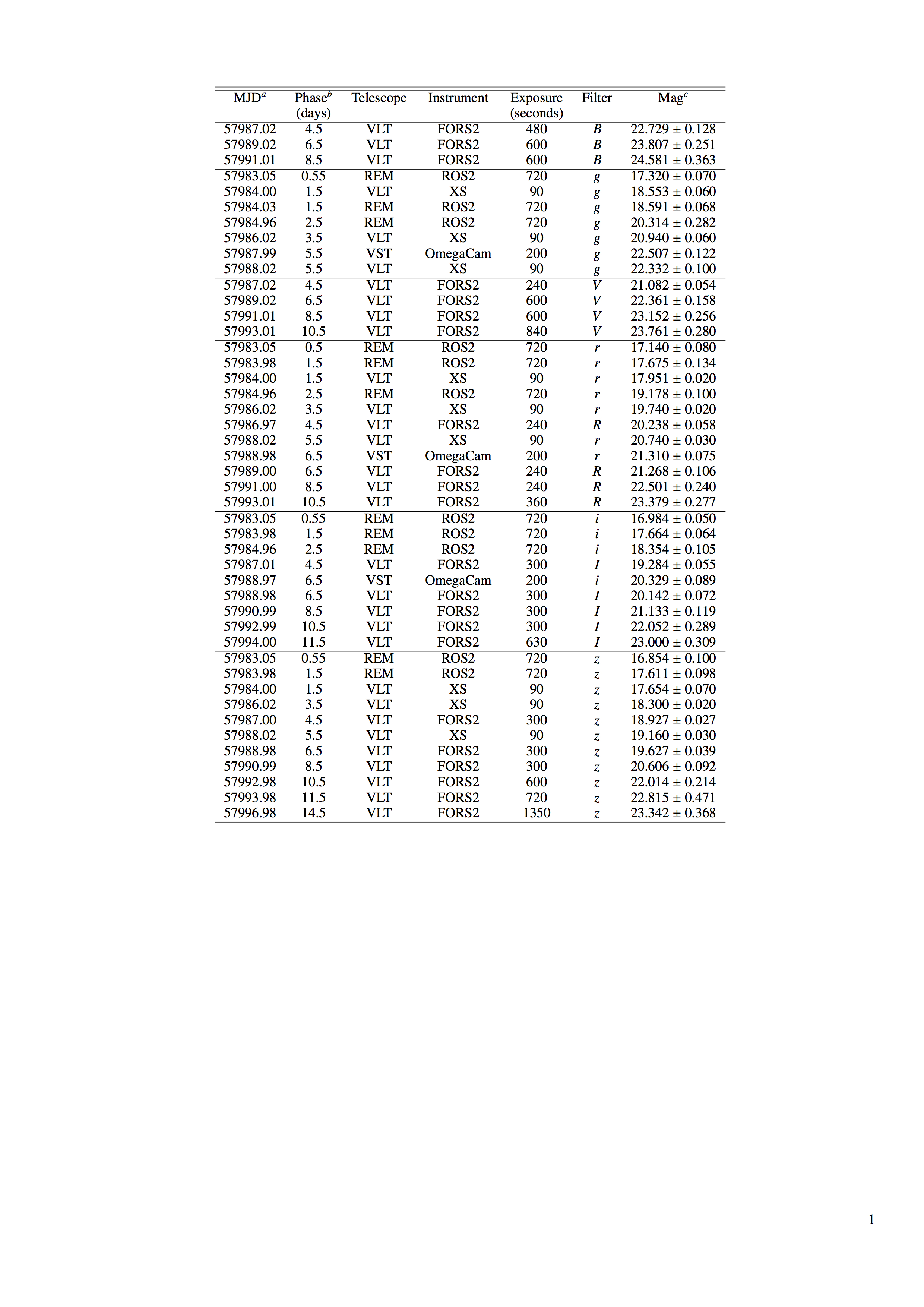}
\end{figure}
\noindent {\bf Extended Data Table 1: Log of photometric observations. } $^a$JD - 2,400,000.5; $^b$After GW trigger time; $^c$AB magnitudes, not corrected for Galactic extinction (E$_{\rm B-V}$=0.11).

%
%
\begin{figure}
	\centering
	\includegraphics[height=18.cm,angle=0]{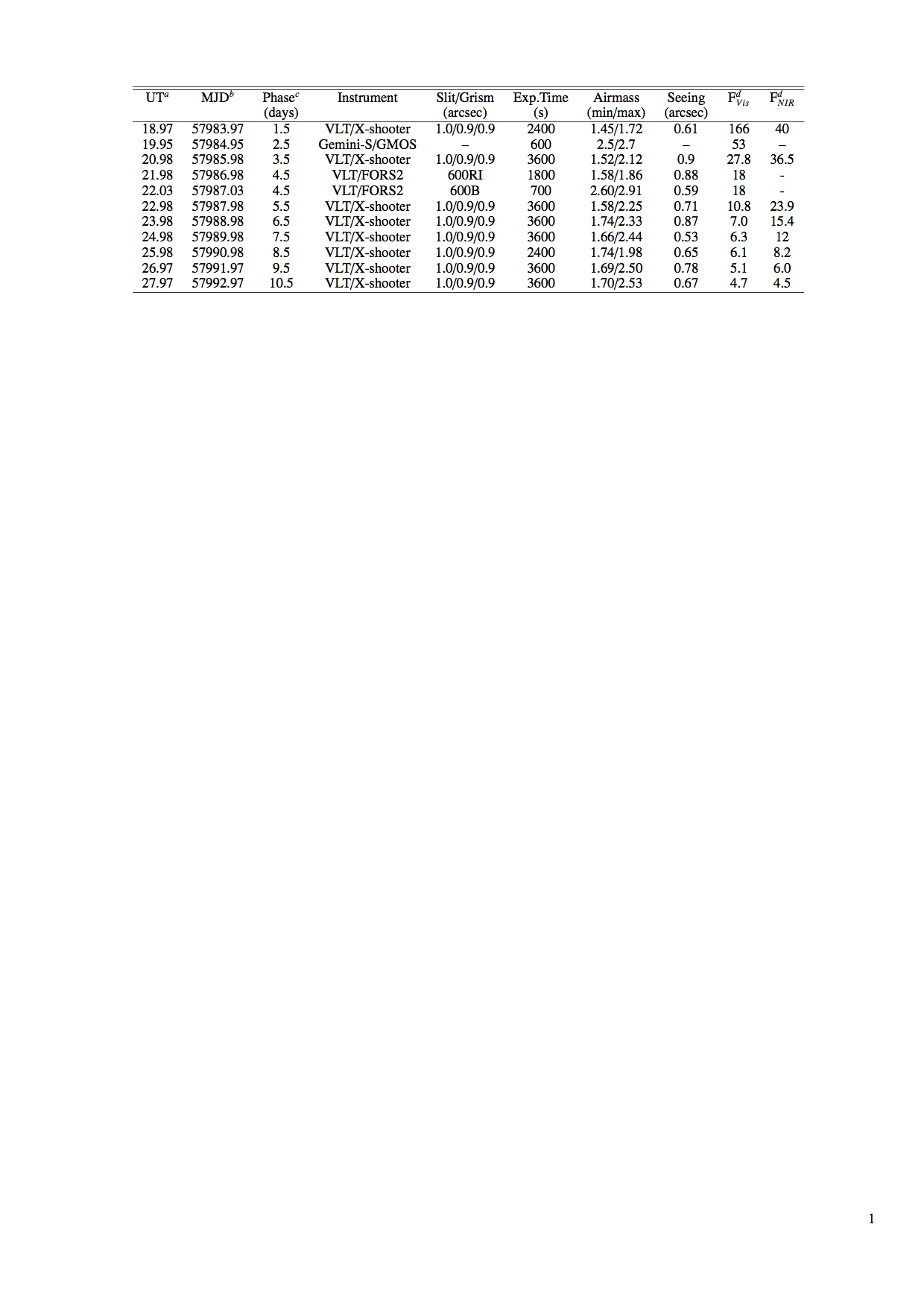}
\end{figure}

\noindent {\bf Extended Data Table 2: Log of spectroscopic observations.}  $^a$ UT days of Aug 2017. $^b$  JD - 2,400,000.5. $^c$ After GW trigger time. $^d$ Fluxes at 6000 and 15000 \AA\ in $10^{-18}$ erg s$^{-1}$ cm$^{-2}$ \AA$^{-1}$, not corrected for reddening; uncertainties are $\sim$10\%.

%
%
\begin{figure}
	\centering
	\includegraphics[height=12.cm,angle=0]{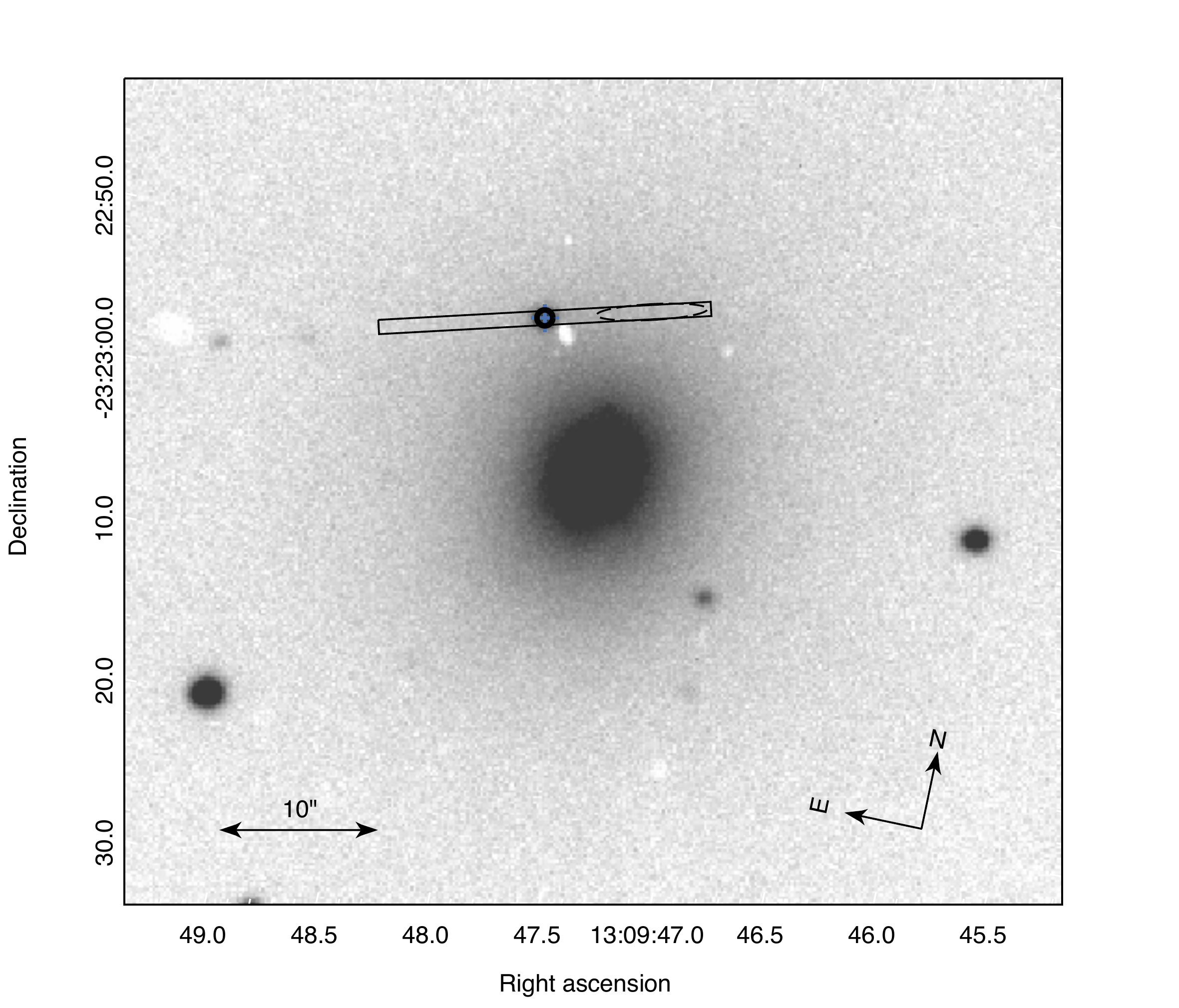}
\end{figure}

\noindent {\bf Extended Data Figure 1: Image of the NGC4993 galaxy}.  The image was obtained with the X-shooter acquisition camera ($z$ filter). The X-shooter slit overlaid in red. The position of the OT has been marked by a blue circle. The position of the line emission in the slit has been also marked. The dust lanes visible in the host intersects the slit at the position of the line emission.

%
%
\begin{figure}
	\centering
	\includegraphics[height=12.cm,angle=0]{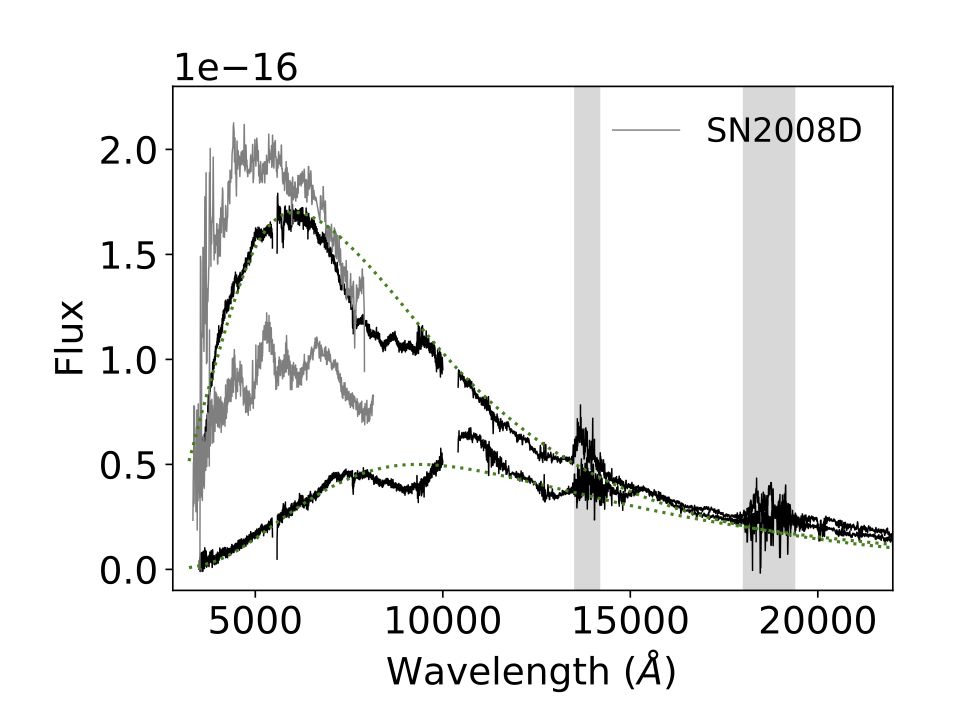}
\end{figure}

\noindent {\bf Extended Data Figure 2: Black-body fit to the SSS17a/DLT17ck spectra}. The two early X-shooter spectra of GW170817, obtained  1.5 and 3.5 d after discovery are compared with  the spectra of the type Ib SN 2008D\cite{mazzali2008} obtained at 2-5 days after explosion respectively (blue, arbitrarily scaled in flux).  The dotted line show the black-body fit of the optical continuum of GW170817 with temperature 5000 and 3200 K respectively.

%
%
\begin{figure}
	\centering
	\includegraphics[height=12.cm,angle=0]{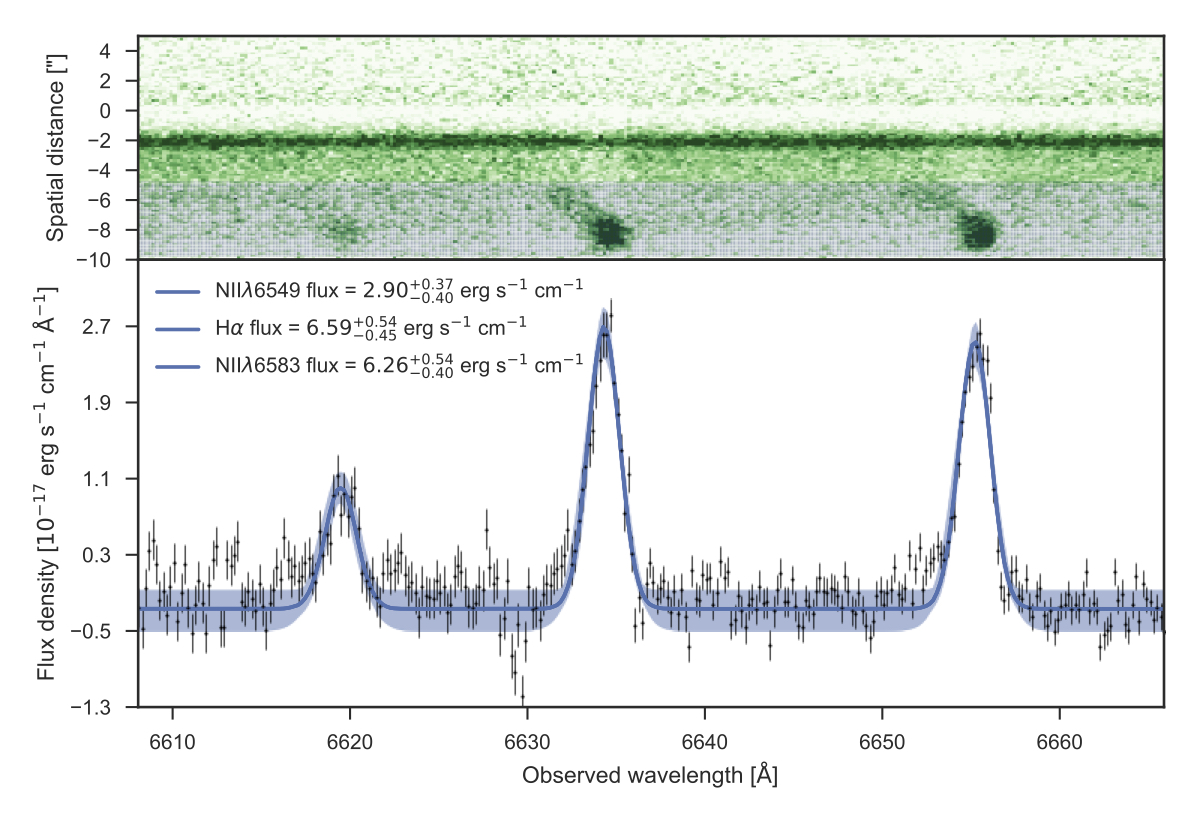}
\end{figure}

\noindent {\bf Extended Data Figure 3: 2D image of the SSS17a/DLT17ck spectrum}. The upper panel shows the rectified, X-shooter 2D-image. The dark line visible across the entire spectral window is the bright continuum of the OT and the offset, dark blobs indicate the position of the line emission from NII$\lambda$6549, H$\alpha$, and NII$\lambda$6583. The lower panel shows an extraction of the line emission where the line fits are overlain. The integrated line fluxes are given in the labels, normalized by a factor of $10^{-17}$ for clarity.

%
%
\begin{figure}
	\centering
	\includegraphics[height=12.cm,angle=0]{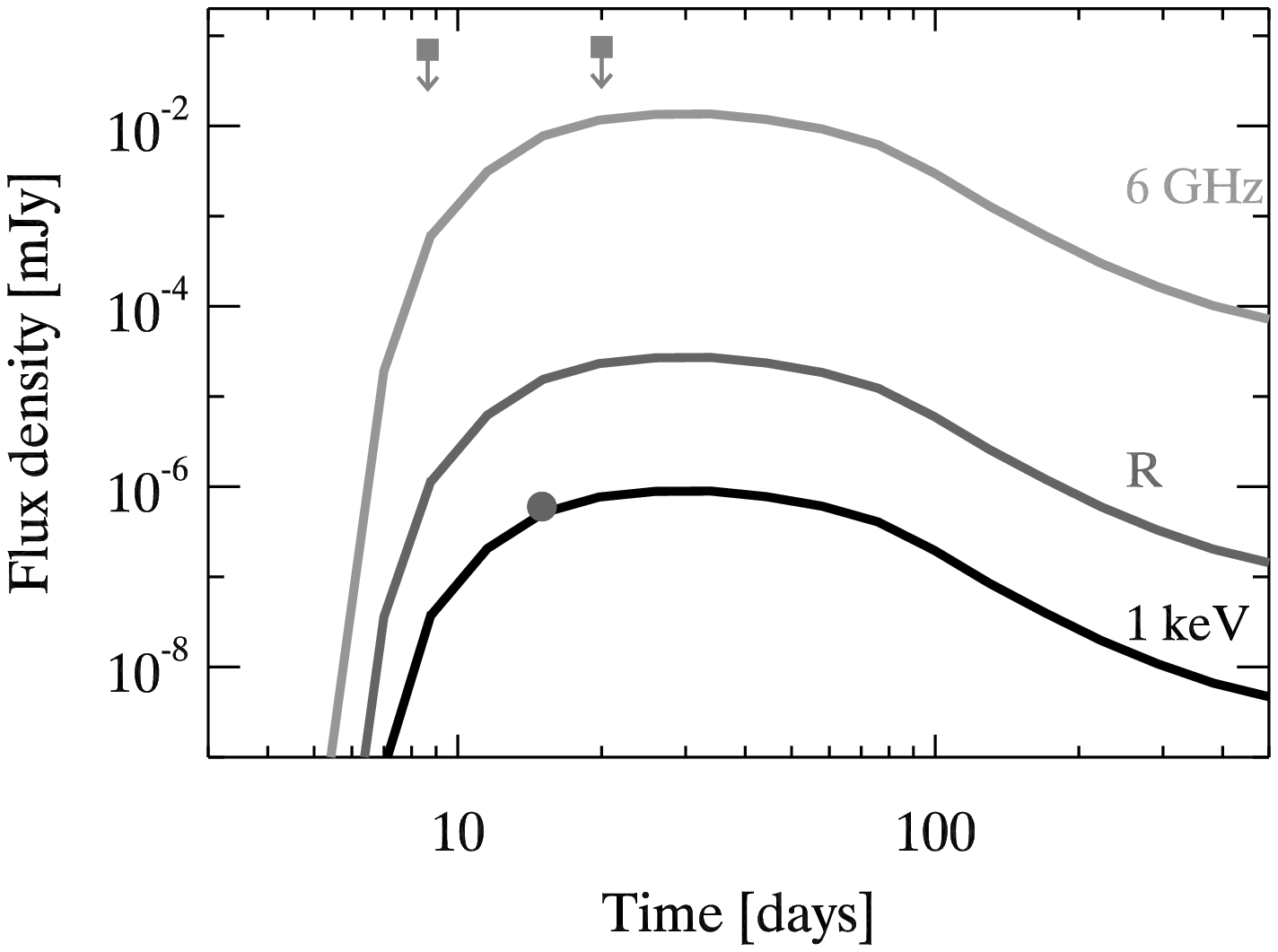}
\end{figure}

\noindent {\bf Extended Data Figure 4: Off-axis GRB afterglow modeling}. Synthetic X-ray, optical and radio light curve of the GRB afterglow as predicted in an off-axis jet model. The filled dot symbol shows the X-ray detection\cite{trojapaper} and the  arrows two representative radio upper limits \cite{Paragi,Mooley}.

\end{document}